\documentclass[multphys,vecphys]{svmult}

\usepackage{makeidx}         
\usepackage{graphicx}        
\usepackage{multicol}        
\usepackage[bottom]{footmisc}

\makeindex             

\begin{document}

\title*{VLTI/MIDI measurements of extended mid-infrared emission in the Galactic Center}
\titlerunning{VLTI/MIDI measurements in the Galactic Center}
\author{J.-U.~Pott\inst{1,2}\and
A.~Eckart\inst{2} \and A.~Glindemann\inst{1} \and T.~Viehmann\inst{2} \and Ch.~Leinert\inst{3}}
\authorrunning{J.-U.~Pott et al.}
\institute{ESO, Karl-Schwarzschildstr.~2, 75478 Garching b.M., Germany
\texttt{jpott@eso.org}
\and I. Physik. Institut, Univ. zu K\"oln, Z\"ulpicher Str.~77, 50937 Cologne, Germany \and Max-Planck-Institut for Astronomy, K\"onigstuhl 17, 69117 Heidelberg, Germany}
%
%
\maketitle
We investigated with MIDI the extension of dusty mid-infrared 
excess sources (IRS~1W, IRS~10W, IRS~2, IRS~8) in immediate vicinity 
to the black hole (BH) at the GC.
We derive 3$\sigma$ upper limits of the correlated fluxes of our target sources which give direct constraints on the size of the emitting regions. Most probably the emission originates from bow shocks generated by windy stars ploughing through the dense matter of the Northern MiniSpiral.
\paragraph{Why is the dust morphology of interest?}
\label{par:1}
At the Center of our Galaxy (GC) star formation (SF) close to a supermassive BH can be studied at unique linear scale.
\newline The existence of massive stars contributing more than 50\% of the ionizing luminosity within the GC confines the latest SF activity to happen at maximum a few Myr ago \cite{Najarro}. 
Simulations show, that a parental molecular cloud can spiral 
into the very center during the lifetime of a massive star, 
providing in particular good conditions for the more massive 
stars. For lower mass stars matter is less bound and will be
easily removed in the vicinity of the massive BH \cite{Gerhard}. 
\newline On the other hand the enigmatic featureless infrared 
excess sources within the MiniSpiral 
(e.g. 1W and 10W) only a few arsec away of SgrA* (1''$\sim$39~mpc) 
could also indicate embedded young stellar objects. Recent near infrared AO imaging suggested 
a different explanation for these sources. 
Tanner et al. (2005) \cite{Tanner} found bow-like morphologies. A more thorough analysis revealed that these bows can be explained by heated dust which is shocked through the interaction of a strong stellar wind 
(v$_\infty \le 1000$ km~s$^{-1}$) from a massive star 
(most favourable Wolf-Rayet type) ploughing through the MiniSpiral.

\paragraph{Observations \& Results}
\label{sec:2}
We conducted a VLTI observation \cite{Pott} in the mid infrared (8-12~$\mu$m) which is totally dominated by thermal re-radiation of the stellar 
UV luminosity by dust. Low resolution dispersed visibility moduli at 20~mas angular scale were obtained in July 2004 using MIDI at the UT2-UT3(47m) baseline. The image motion was corrected with a tip-tilt unit guiding on a 25'' distant optical foreground guide star. 
\newline MIDI provides internal fringe tracking on the basis of group delay fringe tracking. The group delay is estimated by the Fourier transform of the dispersed interferometric spectrum. An optical path difference (OPD) between the two interfering light beams results in a cos-pattern in the interferometric spectrum as long as the OPD is shorter than the coherence length. 
\newline No fringes could be detected on the embedded sources in the MiniSpiral (1W, 2 and 10W). Because the total flux densities of all these sources are above the detection limit, the negative fringe detections showed, that either the VLTI resolved out the entire flux density of the source or
any compact unresolved source component is weaker than the upper limit of the correlated flux density listed in Table~\ref{tab:1}. 
\newline We quantified that statement by giving an upper limit of the correlated flux densities of the different measurements, depending on the actual observing conditions in Table~\ref{tab:1}. Furthermore we assumed for two sources a dust morphology similar to the near infrared findings \cite{Tanner}. Then the lower limit of the width of the bowshock feature can be fitted to the upper limit of the correlated flux. The estimated lower limits of the bowshock widths are well within the near infrared findings. A further constraint on the extension of 1W was achieved by deconvolution of the acquisition image. We could derive an overall extension scale of $\sim$ 350mas which contains the entire bowshock and indicates a significant increase in size of the radiating structures at MIR wavelengths with respect to the near infrared.

\paragraph{Outlook}
\label{sec:4}
The new higher order adaptive optics system MACAO has improved the stability of the beams and concentrates more light into the interferometric field-of-view. The widths found in the NIR are of the order of the lower limits presented here. Therefore positive fringe detection may be within reach already without the external fringe tracker FINITO. 
%
%
\vspace{-0.5cm}
\begin{table}
\centering
\caption{Upper correlated flux limits give lower limits of the source size.}
\label{tab:1}       
%
%
\begin{tabular}{lccc}
\hline\noalign{\smallskip}
Sources with existing bowshock models: &  & IRS~1W & IRS~10W \\
\noalign{\smallskip}\hline\noalign{\smallskip}
Total flux at 8.6micron (VISIR), (extinct. Av=25) & [Jy] &4.6 &2.7 \\
Upper limit of the correlated flux density & [Jy] & $<$~0.3&$<$~0.25\\
Visibility limits&[1]&$<$~0.06&$<$~0.09\\
Width of bowshock models&[mas/AU]&$>$~30/240&$>$~20/160\\
\noalign{\smallskip}\hline
Sources without existing bowshock models:&& IRS~2 & IRS~8  \\
\hline\noalign{\smallskip}
Upper limit of the correlated flux density & [Jy] &$<$~0.4&$<$~0.35\\
\noalign{\smallskip}\hline
\end{tabular}
\end{table}
%
%
%
%
%
%
%
%
\vspace{-1.1cm}
%
%

%
%


\printindex
\end{document}